\documentclass[preprint,amsmath,amssymb,aps,superscriptaddress]{revtex4-2}
\usepackage{amsfonts}
\usepackage{stmaryrd}
\usepackage{bbm}
\usepackage{mathrsfs}
\usepackage{tipa}
\usepackage{amssymb}
\usepackage{txfonts}
\usepackage{graphicx}
\usepackage{dcolumn}
\usepackage{epstopdf}
\usepackage[colorlinks,linkcolor=blue,urlcolor=blue,citecolor=blue]{hyperref}
\usepackage{multirow}
\usepackage{subfigure}
\usepackage{url}
\usepackage{upgreek}
\usepackage[utf8]{inputenc}
\usepackage[english]{babel}

\usepackage{lineno}
\usepackage{xcolor}
\usepackage{ulem}

\begin{document}

\title{Robust Antiferromagnetic Ordering above 500 K in a Family of Bilayer van der Waals Materials}

\author{Tengda Rong}
\affiliation{Institute of Neutron Science and Technology, Guangdong Provincial Key Laboratory of Magnetoelectric Physics and Devices,\\School of Physics, Sun Yat-Sen University, Guangzhou, Guangdong 510275, China}

\author{Linxin Wu}
\affiliation{Institute of Neutron Science and Technology, Guangdong Provincial Key Laboratory of Magnetoelectric Physics and Devices,\\School of Physics, Sun Yat-Sen University, Guangzhou, Guangdong 510275, China}

\author{Xingke Fu}
\affiliation{Beijing National Laboratory for Condensed Matter Physics, Institute of Physics, Chinese Academy of Sciences, Beijing, China}

\author{Anqi Huang}
\affiliation{Institute of Neutron Science and Technology, Guangdong Provincial Key Laboratory of Magnetoelectric Physics and Devices,\\School of Physics, Sun Yat-Sen University, Guangzhou, Guangdong 510275, China}

\author{Feiran Shen}
\affiliation{Spallation Neutron Source Science Center, Dongguan, Guangdong, China}

\author{Zixian Dong}
\affiliation{Institute of Neutron Science and Technology, Guangdong Provincial Key Laboratory of Magnetoelectric Physics and Devices,\\School of Physics, Sun Yat-Sen University, Guangzhou, Guangdong 510275, China}

\author{Jianwei Huang}
\affiliation{Institute of Neutron Science and Technology, Guangdong Provincial Key Laboratory of Magnetoelectric Physics and Devices,\\School of Physics, Sun Yat-Sen University, Guangzhou, Guangdong 510275, China}

\author{Hao Li}
\affiliation{Institute of Nuclear Physics and Chemistry, Chinese Academy of Engineering Physics, Mianyang, Sichuan, China}
\affiliation{National Key Laboratory for Neutron Science and Technology, Mianyang, Sichuan, China}

\author{Benyuan Cheng}
\affiliation{Shanghai Institute of Laser Plasma, Shanghai, China}
\affiliation{Center for High Pressure Science and Technology Advanced Research, Shanghai, China}

\author{Lunhua He}
\affiliation{Spallation Neutron Source Science Center, Dongguan, Guangdong, China}
\affiliation{Beijing National Laboratory for Condensed Matter Physics, Institute of Physics, Chinese Academy of Sciences, Beijing, China}

\author{Long Jiang}
\affiliation{Instrumental Analysis and Research Center, Sun Yat-Sen University, Guangzhou, Guangdong, China}

\author{Kun Cao}
\affiliation{Institute of Neutron Science and Technology, Guangdong Provincial Key Laboratory of Magnetoelectric Physics and Devices,\\School of Physics, Sun Yat-Sen University, Guangzhou, Guangdong 510275, China}

\author{Zhen Chen}
\affiliation{Beijing National Laboratory for Condensed Matter Physics, Institute of Physics, Chinese Academy of Sciences, Beijing, China}
\affiliation{School of Physical Sciences, University of Chinese Academy of Sciences, Beijing, China}

\author{Meng Wang}
\email{wangmeng5@mail.sysu.edu.cn}
\affiliation{Institute of Neutron Science and Technology, Guangdong Provincial Key Laboratory of Magnetoelectric Physics and Devices,\\School of Physics, Sun Yat-Sen University, Guangzhou, Guangdong 510275, China}

\author{Xiang Chen}
\email{chenx889@mail.sysu.edu.cn}
\affiliation{Institute of Neutron Science and Technology, Guangdong Provincial Key Laboratory of Magnetoelectric Physics and Devices,\\School of Physics, Sun Yat-Sen University, Guangzhou, Guangdong 510275, China}

\date{\today}

\begin{abstract}
Stabilizing magnetic order above room temperature in layered materials is highly desirable for low-dimensional spintronics applications. However, realizing room-temperature antiferromagnetism in layered magnets remains challenging. Here, we present a layered semiconducting oxyhalide material family, Sr$_3$Fe$_2$O$_5$\textit{X}$_2$ (\textit{X} = F, Cl, Br), which exhibits robust high-temperature antiferromagnetic ordering. The two-dimensional nature and scalability of this system are demonstrated by atomically resolving the van der Waals gap of Sr$_3$Fe$_2$O$_5$Br$_2$ and synthesizing centimeter-scale single crystals of Sr$_3$Fe$_2$O$_5$Cl$_2$. Temperature-dependent neutron scattering and magnetization measurements determine their G-type magnetic structure and transition temperature. Remarkably, the N\'eel temperature of Sr$_3$Fe$_2$O$_5$F$_2$ ($\sim$580 K) is the highest among all reported layered antiferromagnets.Density functional theory calculations demonstrate that strong nearest-neighbor superexchange interactions along both out-of-plane and in-plane directions, particularly the magnetic couplings along lattice $c$ direction within the bilayer motif, are primarily responsible for the high N\'eel temperatures in Sr$_3$Fe$_2$O$_5$\textit{X}$_2$. In contrast, inter-bilayer spin couplings are negligibly weak, without much influence on the stabilization of the antiferromagnetic order. Therefore, theoretical calculations predict that this robust antiferromagnetic order could persist or even strengthens at room-temperature even at the monolayer limit. Our results highlight iron-based layered oxyhalides as a promising platform for exploring beyond-room-temperature antiferromagnetism in the two-dimensional limit, holding immense potential in areas like information storage and spin-based devices.

\end{abstract}

\maketitle 
\clearpage

\section{Introduction}
 Magnetism in van der Waals (vdW) materials with dimensional confinement has emerged as a prominent frontier in condensed matter physics \cite{J.Phys.Condens.Matter.(2016)park, Nature(2018)Burch, Nat.Nanotechnol(2019)Gibertini, Science(2019)Gong, Annalen.der.Physi(2020)Wang,RMP(2026)Park}. Intrinsically weak interlayer couplings, because of the loose bondings by van der Waals force, facilitate easier isolation of monolayer and few-layer flakes via mechanical exfoliation. Such pronounced dimensional confinement provides an excellent platform for investigating unconventional physical properties and emergent phenomena at the nanoscale. The ability to arbitrarily stack these layers offers further remarkable flexibility in materials design, stimulating extensive research into engineered heterostructures and superlattices ~\cite{Nat.Nanotechnol(2019)Gibertini, Science(2019)Gong}. Exceptional phenomena include unconventional superconductivity and fractional quantum anomalous Hall states are observed in twisted superlattices ~\cite{Nature(2018)cao, Nature(2023)Park}. The discovery of intrinsic magnetic order in atomically thin crystals ~\cite{NanoLetters(2016)Lee, Nature(2017)Huang, Nature(2017)Gong} has further promoted device research, leading to advances in spin field-effect transistors ~\cite{NanoLetters(2016)Lee}, quantum light emitters ~\cite{Nat.Mat(2023)Li}, quantum computing ~\cite{ACSnano(2025)Lee,NC(2024)Xiong}, etc.

A critical limitation, however, is that the magnetic transition temperatures of most vdW magnets are well below room temperature, hindering practical applications. To address this, considerable efforts have been devoted to elevating the magnetic ordering temperatures in recent years, with significant progress reported in ferromagnets such as $\mathrm{Fe_5GeTe_2}$~\cite{ACSnano(2019)MayAF, PRB(2020)Zhang}, Ni- and Co- doped $\mathrm{Fe_5GeTe_2}$ \cite{PRL(2022)Chen, PRM(2020)AFMay, APL(2020)Tian, Sci.adv(2022)zhangHongrui}, $\mathrm{Fe_3GaTe_2}$ \cite{NC(2022)zhang, NC(2024)zhangHongrui}, etc. For antiferromagnetic vdW magnets, materials with N\'eel temperature ($T_{\mathrm{N}}$) reaching room temperature are still quite rare. A-type antiferromagnetic order with $T_{\mathrm{N}} \sim 340$~K was confirmed in 40\% Co doped $\mathrm{Fe_5GeTe_2}$ \cite{PRM(2020)AFMay, PRM(2022)Chen}. This compound is metallic with in-plane ferromagnetism and is antiferromagnetically coupled along the out-of-plane direction, rendering a magnetic propagation wavevector $\boldsymbol{k} = (0, 0, 1/2)$  \cite{PRM(2022)Chen}. Still, vdW antiferromagnets with non-trivial spin textures and an insulating electronic ground state at or above room temperature are highly desired.

A promising strategy involves incorporating multiple anion species into the crystal lattice. The significant disparities in ionic radius, charge, and electronegativity between various anions expand the degrees of freedom for modulating the atomic and electronic structures of materials ~\cite{NC(2018)kageyama}. This mixed-anion strategy is particularly effective in Ruddlesden-Popper phase (RP-phase) oxyhalides. In these materials, halide atoms (F, Cl, Br, I) tend to occupy certain ordered oxygen sites, segregating oxygen and halide atoms into distinct layers. Such a unique configuration often forms low-dimensional layered structures, allowing control over the dimensionality, electronic properties and spin exchange of magnetic interactions ~\cite{NC(2018)kageyama}. This principle is exemplified in superconducting compounds such as $\mathrm{Sr_2CuO_2F_{2+\delta}}$ ~\cite{Nature(1994)Mamour} and $\mathrm{(Ca,Na)_2CuO_2Cl_2}$ ~\cite{Nature(1994)Hiroi}, as well as geometrically frustrated $\mathrm{(CuCl)LaNd_2O_7}$ ~\cite{PRB(2009)Tsirlin}. Notably, many RP-phase transition metal ion based oxyhalides exhibit high-temperature magnetic ordering ~\cite{JMC(2001)Knee}. Particularly, the iron-based oxyhalides $\mathrm{Sr_3Fe_2O_5Br_2}$ and $\mathrm{Sr_3Fe_2O_5Cl_2}$ serve as prototypical candidates within this extensive family, in which N\'eel temperatures above 500~K were reported~\cite{Solid.State.Sci.(2004)Knee, PRM(2025)May.AF}.

In this work, we optimize the systematic synthesis of the layered oxyhalide family $\mathrm{Sr_3Fe_2O_5\textit{X}_2}$ (\textit{X} = F, Cl, Br) (Figure~\ref{fig1}(a)), including the synthesis of both powder and single crystal sample forms. The largest lateral dimension of $\mathrm{Sr_3Fe_2O_5Cl_2}$ single crystals can reach more than 20~mm. The phase purity, as well as the layered structure of $\mathrm{Sr_3Fe_2O_5\textit{X}_2}$, are confirmed by X-ray diffraction (XRD) and neutron powder diffraction (NPD). In addition, ultrathin flakes of $\mathrm{Sr_3Fe_2O_5Br_2}$ are readily isolated via mechanical exfoliation down to the thickness of one-unit-cell ($\sim 3$~nm), and the intrinsic van der Waals spatial gaps are directly resolved at the atomic scale by scanning transmission electron microscopy (STEM). Magnetization measurements indicate AFM ground state in $\mathrm{Sr_3Fe_2O_5\textit{X}_2}$, with N\'eel temperature decreasing from the largest $T_{\mathrm{N}} \sim 580$~K for \textit{X} = F, to $\sim 540$~K for both \textit{X} = Cl and Br. NPD studies reveal a G-type magnetic structure with in-plane spins and a magnetic propagation wavevector $\boldsymbol{k} = (1/2, 1/2, 0)$ in this material family. Theoretical calculations reveal that the robust high-temperature magnetic order in this system is primarily driven by strong out-of-plane and in-plane nearest-neighbor superexchange couplings. This long-range magnetic order is predicted to persist down to the two-dimensional (2D) limit without reducing its N\'eel temperature. Lastly, both UV-Vis diffuse reflectance spectroscopy (DRS) and angle-resolved photoemission spectroscopy (ARPES) studies indicate that all three compounds are semiconductors with both optical and electronic bandgaps larger than 2~eV. Our work highlights that $\mathrm{Sr_3Fe_2O_5\textit{X}_2}$ (\textit{X} = F, Cl, Br) is a promising layered semiconducting material platform with N\'eel temperature approaching 600~K, and might still hold its N\'eel temperatures even down to the 2D limit.

\section{Results and discussion}
\subsection{Sample synthesis and Crystal structure}\label{subsec2}
Although some synthesis was reported before~\cite{J.Solid.State.Chem.(1991)Ackerman, PRM(2025)May.AF}, we further optimize the sample growth methods and extend to the whole $\mathrm{Sr_3Fe_2O_5\textit{X}_2}$ (\textit{X} = F, Cl, Br) family. Both polycrystalline $\mathrm{Sr_3Fe_2O_5Cl_2}$ and $\mathrm{Sr_3Fe_2O_5Br_2}$ can be synthesized via conventional solid-state synthesis, while polycrystalline $\mathrm{Sr_3Fe_2O_5F_2}$ is obtained via chemical fluorination of the precursor $\mathrm{Sr_3Fe_2O_7}$ powders. The ratio of poly (vinylidene fluoride) (PVDF) to $\mathrm{Sr_3Fe_2O_7}$ is very critical and needs to be fine-tuned, in order to obtain a phase-pure, layered form of $\mathrm{Sr_3Fe_2O_5F_2}$. As for the single-crystalline form of $\mathrm{Sr_3Fe_2O_5\textit{X}_2}$, $\mathrm{Sr_3Fe_2O_5Cl_2}$ crystals can be readily obtained via the widely used $\mathrm{SrCl_2}$ flux method. We further optimize the flux growth parameters for $\mathrm{Sr_3Fe_2O_5Cl_2}$, producing large single crystals with lateral dimensions reaching $\sim 20$~mm (Figure~\ref{fig2}(a)). Similar method applied to the growth of $\mathrm{Sr_3Fe_2O_5Br_2}$ single crystals, however, only yields small (sub-millimeter scale) and brittle thin flakes. To obtain large single crystals of $\mathrm{Sr_3Fe_2O_5Br_2}$, a new flux growth involves $\mathrm{FeC_2O_6H_4}$ instead of $\mathrm{Fe_2O_3}$ in a Czochralski furnace with a temperature gradient is utilized, producing large single crystals with lateral dimensions reaching $\sim 5$~mm (Figure~\ref{fig2}(a)).

In order to confirm the layered crystal structure and phase purity of $\mathrm{Sr_3Fe_2O_5\textit{X}_2}$, diffraction and imaging measurements were performed (Figures~\ref{fig1}, ~\ref{fig2}, S1). First, powder XRD measurements on all three polycrystalline samples were collected at room temperature. The XRD patterns can be indexed with a tetragonal structure with a space group $I4/mmm$, as shown in Figures~\ref{fig2}(b), ~\ref{fig2}(d), and~\ref{fig2}(f). From Rietveld refinements, all halide atoms occupy the outer apical oxygen sites (Figure~\ref{fig1}(b)), because of which the neighboring planes of halides are bonded through van der Waals interactions, leading to a layered structure with vdW gaps and micaceous character. The plate-like single crystals of $\mathrm{Sr_3Fe_2O_5Cl_2}$ and $\mathrm{Sr_3Fe_2O_5Br_2}$ are also characterized by XRD, revealing only sharp (0 0 $L$) Bragg reflections (Figures~\ref{fig2}(c) and~\ref{fig2}(e)). To further confirm the crystalline quality of $\mathrm{Sr_3Fe_2O_5Br_2}$, single crystal XRD measurements are performed at 150~K (Figure S1). No impurity phase was detected and the $I4/mmm$ tetragonal structure is verified as well.

Figures~\ref{fig2}(g)-(i) and Table~\ref{tab1} summarize the evolution of the obtained lattice parameters and some select bond lengths and angles from powder XRD refinements, as a function of the halide species \textit{X}. Upon halide substitution from \textit{X} = F to Br, both the lattice parameters $a$ and $c$ increase monotonically. However, this expansion is significantly more pronounced along the lattice $c$ direction than within the $ab$-plane. Quantitatively, from \textit{X} = F, to Cl and Br, lattice parameter $a$ increases from 3.91~\AA{} to 3.95~\AA{} and 3.96~\AA{}, which displays an increase of only 0.05~\AA{} (about 1\% enhancement from \textit{X} = F to \textit{X} = Br); while lattice parameter $c$ increases from 21.38~\AA{} to 23.80~\AA{} and eventually to 25.42~\AA{}, marking an enormous elongation of 4.04~\AA{} (about 19\% increase from \textit{X} = F to \textit{X} = Br). This anisotropic lattice expansion primarily originates from the increase in the ionic radius of the constituent halide ions and their electronegativity, where the halide ions only occupy the apical oxygen sites. Both a larger ionic radius and smaller electronegativity favor a longer bond length, as exemplified in the increase of the Fe-\textit{X} bond distances (Figure~\ref{fig2}(h)). This is directly related to the increase of lattice parameter $c$, and the significant enlargement of the vdW spatial gaps as well, from 1.65~\AA{}, to 2.13~\AA{}, and eventually to 2.48~\AA{}, for \textit{X} = F, Cl, Br, respectively (Figure~\ref{fig2}(h)). It should be noted that, within the bilayer motif, the Fe-O1-Fe bond length $d_{\mathrm{Fe-O1-Fe}}$ and O2-Fe-O2 angle $\theta_{\mathrm{O2-Fe-O2}}$ do not evolve monotonically, as shown in Figure~\ref{fig2}(i).

\subsection{Scanning transmission electron microscopy (STEM)}\label{subsec2}
\begin{figure*}[h]
\centering
\includegraphics[width=0.8\textwidth]{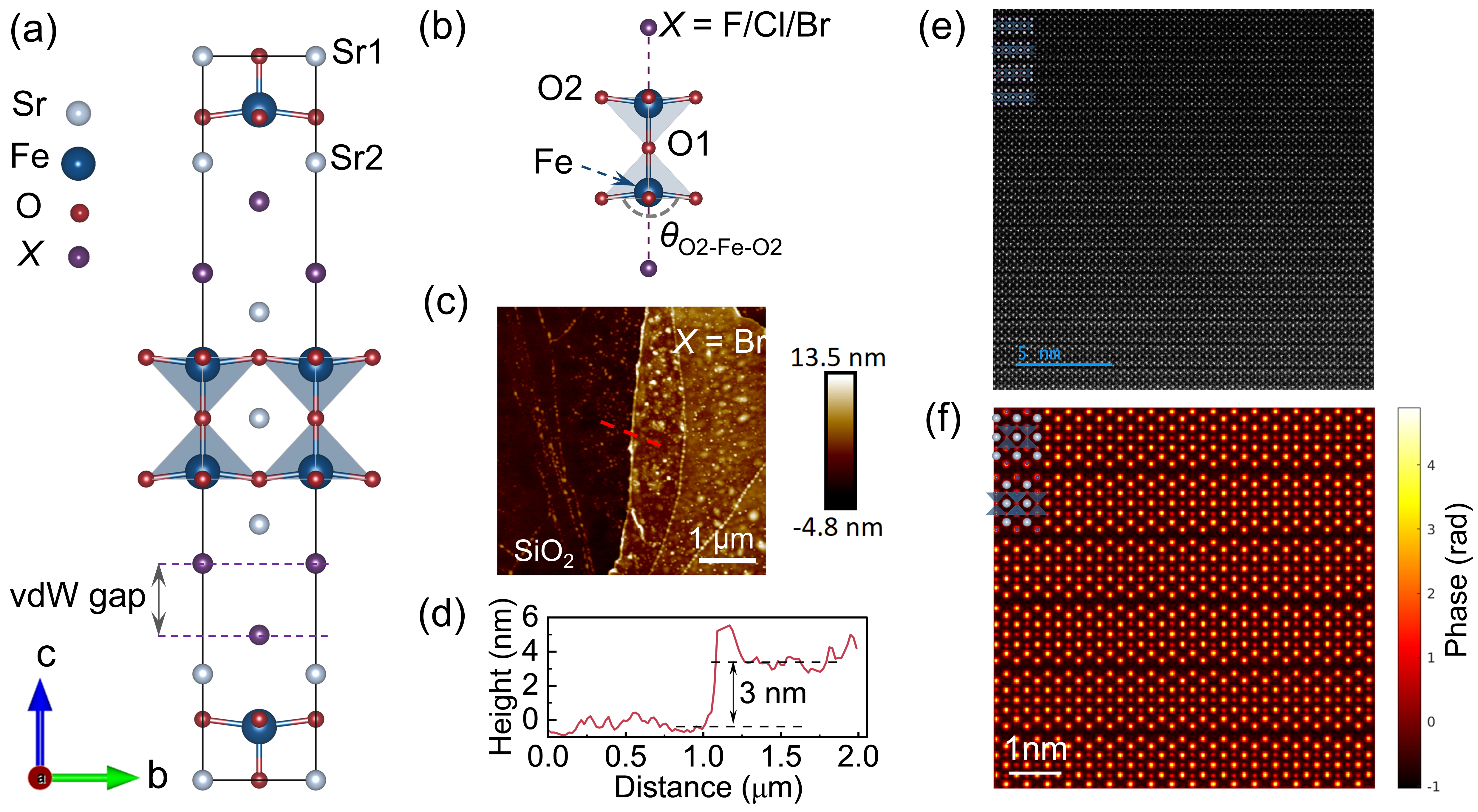}
\caption{Layered structure of $\mathrm{Sr_3Fe_2O_5\textit{X}_2}$ (\textit{X} = F, Cl, Br). (a) Crystal structure of $\mathrm{Sr_3Fe_2O_5\textit{X}_2}$, projected along the [1 0 0] crystallographic axis direction. (b) Bilayer structure with two corner sharing $\mathrm{FeO_5\textit{X}}$ (\textit{X}=F/Cl/Br) octahedrons. (c) Atomic force microscopy image of the $\mathrm{Sr_3Fe_2O_5Br_2}$ (SFOB) flake exfoliated onto a $\mathrm{SiO_2}$ substrate. (d) Height profile along the line cut in (c), indicating a thickness of $\sim 3$~nm, which is about one unit cell of lattice $c$. (e)-(f) High-angle annular dark-field (HAADF) image and projected multislice electron ptychography (MEP) reconstruction, along the [100] zone axis of single-crystal $\mathrm{Sr_3Fe_2O_5Br_2}$, respectively.}
\label{fig1}
\end{figure*}

\begin{figure*}[t]
\centering
\includegraphics[width=0.8\textwidth]{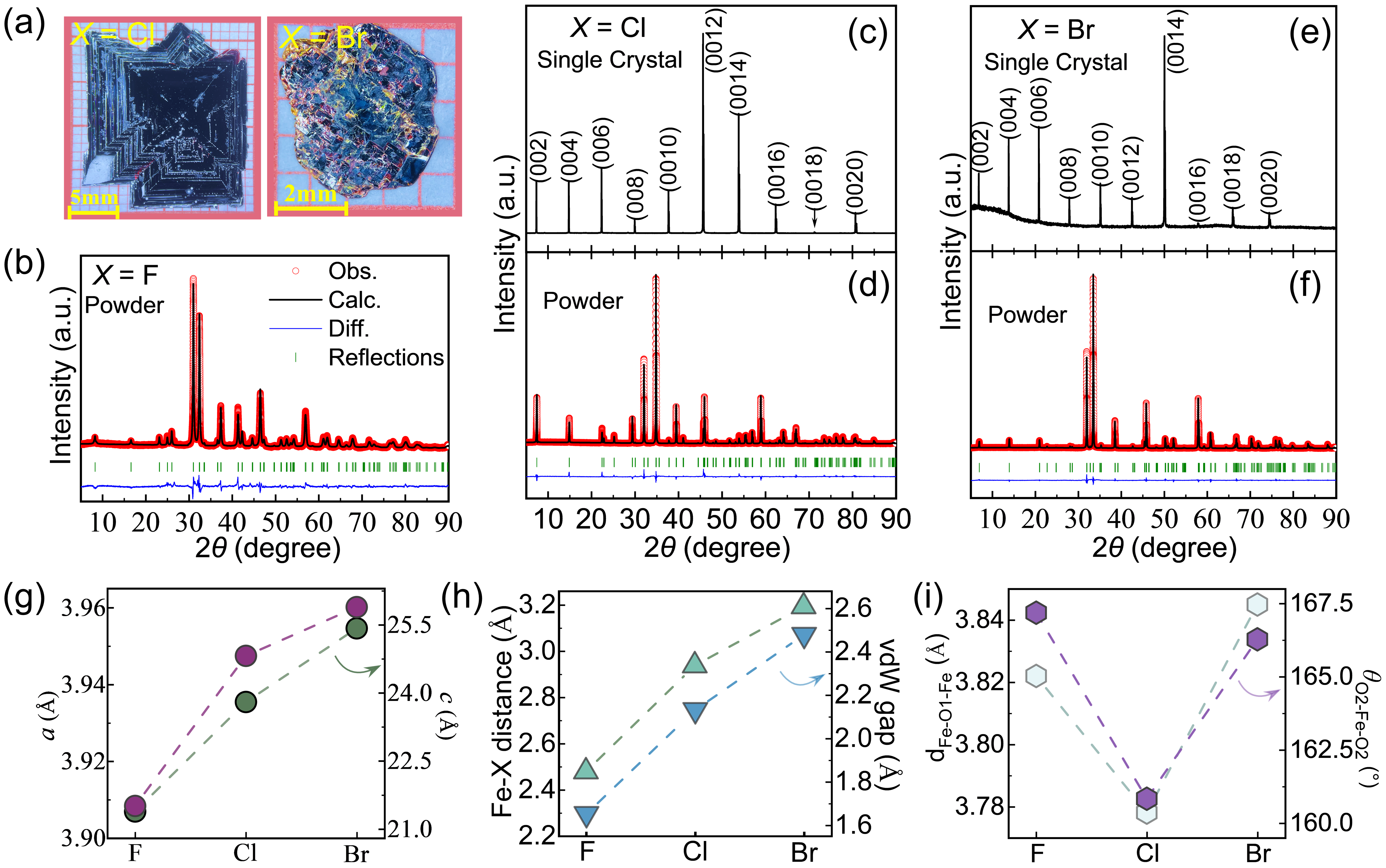}
\caption{X-ray diffraction (XRD) patterns and refined structural parameters of $\mathrm{Sr_3Fe_2O_5\textit{X}_2}$ (\textit{X} = F, Cl, Br) powders at room temperature. (a) Optical micrographs of the $\mathrm{Sr_3Fe_2O_5Cl_2}$ and $\mathrm{Sr_3Fe_2O_5Br_2}$ single crystals. (b)(d)(f) Powder XRD pattern of polycrystalline $\mathrm{Sr_3Fe_2O_5\textit{X}_2}$ at room temperature: (b) \textit{X} = F, (d) \textit{X} = Cl and (f) \textit{X} = Br. The panels of (c) and (e) indicate the XRD patterns of $\mathrm{Sr_3Fe_2O_5Cl_2}$ and $\mathrm{Sr_3Fe_2O_5Br_2}$ single crystals along the (0 0 $L$) direction. (g)-(i) Select structural information from powder XRD refinements of $\mathrm{Sr_3Fe_2O_5\textit{X}_2}$: (g) lattice parameters $a$ and $c$, (h) Fe-\textit{X} bond lengths and van der Waals spatial gaps, (i) $\mathrm{Fe-O_1-Fe}$ bond lengths $d_{\mathrm{Fe-O1-Fe}}$ and $\mathrm{O_2-Fe-O_2}$ bond angles $\theta_{\mathrm{O2-Fe-O2}}$.}
\label{fig2}
\end{figure*}

Ultrathin nanosheets of $\mathrm{Sr_3Fe_2O_5Br_2}$ were isolated via mechanical exfoliation, with a typical optical micrograph shown in Figure S2(a). Atomic force microscopy was used to determine the thickness of the flakes. The atomic force microscopy measurements reveal a topographic height of approximately 3~nm (Figures~\ref{fig1}(c) and~\ref{fig1}(d)). This value is close to the value of lattice parameter $c$, which corresponds to one unit cell (u.c.) thickness. To directly observe the van der Waals spatial gap at the atomic scale, STEM was used to further characterize the local structure of $\mathrm{Sr_3Fe_2O_5Br_2}$ single crystals. Figures~\ref{fig1}(e) and~\ref{fig1}(f) display a high-angle annular dark-field (HAADF) image and a multislice electron ptychography (MEP) image acquired along the [100] zone axis, respectively. The characteristic bilayer RP-phase structure is clearly resolved. Specifically, the apical sites are exclusively occupied by Br atoms across the entire field of view. This uniform occupation confirms the presence of long-range structural order and establishes a distinct van der Waals gap between adjacent Br atomic layers. These findings indicate the easily cleavable nature of the vdW materials $\mathrm{Sr_3Fe_2O_5\textit{X}_2}$, particularly for \textit{X} = Br, where the interlayer spatial gap is largest (about 2.48~\AA{}).

\subsection{Magnetic properties}\label{subsec2}
\begin{figure*}[t]
\centering
\includegraphics[width=0.8\textwidth]{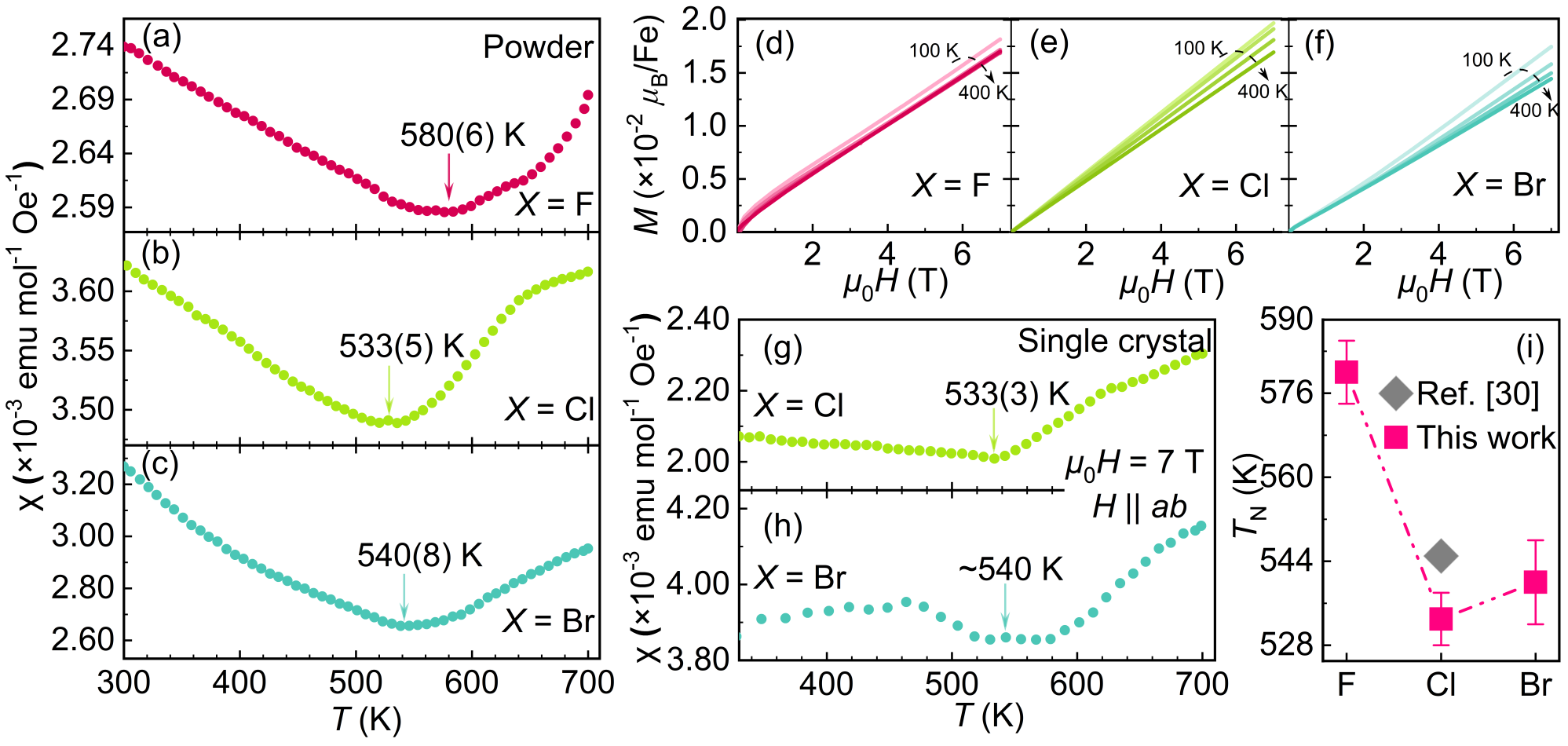}
\caption{Magnetization data of $\mathrm{Sr_3Fe_2O_5\textit{X}_2}$. (a)-(c) Temperature-dependent magnetization of the three polycrystalline samples measured under an applied magnetic field of 7~T. (d)-(f) Isothermal magnetization of $\mathrm{Sr_3Fe_2O_5\textit{X}_2}$ measured from 100~K to 400~K (step 100~K). Temperature-dependent magnetization of (g) $\mathrm{Sr_3Fe_2O_5Cl_2}$ and (h) $\mathrm{Sr_3Fe_2O_5Br_2}$ single crystals measured with the magnetic field applied parallel to the $ab$-plane. (i) The magnetic transition temperature as a function of halogen \textit{X} substitution. Solid diamond denotes the $T_{\mathrm{N}}$ of $\mathrm{Sr_3Fe_2O_5Cl_2}$ determined by M\"ossbauer spectroscopy~\cite{PRM(2025)May.AF}.}
\label{fig3}
\end{figure*}

Figure~\ref{fig3} shows the magnetization data for all synthesized $\mathrm{Sr_3Fe_2O_5\textit{X}_2}$ compounds. The temperature dependent susceptibility of the polycrystalline $\mathrm{Sr_3Fe_2O_5\textit{X}_2}$ sample is presented in Figures~\ref{fig3}(a)-(c). Owing to strong antiferromagnetic interactions, the susceptibility data exhibit only weak temperature dependence and display smooth minima at approximately 580~K, 533~K and 540~K for \textit{X} = F, Cl, Br, respectively. These minima of the susceptibility data correspond to their N\'eel temperatures. For instance, the N\'eel temperature $T_{\mathrm{N}} \sim 545$~K of $\mathrm{Sr_3Fe_2O_5Cl_2}$ is determined from M\"{o}ssbauer measurements ~\cite{PRM(2025)May.AF}, fully consistent with our susceptibility data. Above the AFM transition temperature, the susceptibility still increases, likely due to the presence of strong short-range correlations that are progressively suppressed towards even higher temperatures above $T_{\mathrm{N}}$. Isothermal magnetization measurements on these polycrystalline samples reveal linear behavior at select temperatures from 100~K to 400~K (Figures~\ref{fig3}(d)-(f)), directly confirming the AFM nature of $\mathrm{Sr_3Fe_2O_5\textit{X}_2}$ at room temperature. We also performed temperature dependent susceptibility study on the $\mathrm{Sr_3Fe_2O_5Cl_2}$ and $\mathrm{Sr_3Fe_2O_5Br_2}$ single crystals with a 7~T magnetic field applied along the $ab$-plane direction, demonstrating similar characteristics to their polycrystalline counterparts (Figures~\ref{fig3}(g)-(h)). Combining all magnetization data together, it is evident that the layered oxyhalides $\mathrm{Sr_3Fe_2O_5\textit{X}_2}$ exhibit robust beyond room temperature antiferromagnetism.

Figure~\ref{fig3}(i) summarizes the N\'eel temperature as a function of the halide species \textit{X} in $\mathrm{Sr_3Fe_2O_5\textit{X}_2}$. The magnetic transition temperature shows a decreasing trend from \textit{X} = F to Br, Cl. In particular, $\mathrm{Sr_3Fe_2O_5F_2}$ has the largest $T_{\mathrm{N}} \sim 580$~K among all reported, layered vdW antiferromagnets, as far as we are concerned. It should be noted that because of the much smaller ionic radius of $\mathrm{F^-}$, oxyfluorides tend to possess a significant amount of disorder. For example, $\mathrm{Sr_3Fe_2O_{5.45}F_{1.56}}$ powder was synthesized before, but it exhibits pronounced static anion disorder. As a result, its reported $T_{\mathrm{N}}$ is only $\sim 390$~K~\cite{Chem.Mat(2011)Tsujimoto}, markedly lower than $T_{\mathrm{N}} \sim 580$~K in our work. As a further comparison, among single-layer RP oxyhalides, disordered oxyfluorides such as $\mathrm{Sr_2CoO_3F}$ ($T_{\mathrm{N}} \approx 323$~K)~\cite{Ino.Chem(2012)Tsujimoto} and $\mathrm{Sr_2NiO_3F}$ ($T_{\mathrm{N}} \approx 18$~K)~\cite{PRM(2022)Tsujimoto} all exhibit lower N\'eel temperatures than their Cl-ordered counterparts $\mathrm{Sr_2CoO_3Cl}$ ($T_{\mathrm{N}} \approx 330$~K)~\cite{PRB(2003)knee} and $\mathrm{Sr_2NiO_3Cl}$ ($T_{\mathrm{N}} \approx 28$~K)~\cite{PRM(2022)Tsujimoto, Ino.Chem(2013)Tsujimoto}. Furthermore, $\mathrm{Sr_2FeO_3F}$ ($S = 5/2$) with a low disorder level ($\sim 5\%$) exhibits a transition temperature of approximately 340~K, approaching that of the fully ordered phase ($\sim 358$~K)~\cite{Cry.Gro.&Des(2014)Tsujimoto}. Collectively, these results suggest that the N\'eel temperature of $\mathrm{Sr_3Fe_2O_5F_2}$ is very sensitive to the degree of F-ion disorder, highlighting the importance and necessity to realize a fully ordered phase in $\mathrm{Sr_3Fe_2O_5F_2}$ and other vdW magnets.

\begin{figure*}[t]
\centering
\includegraphics[width=0.8\textwidth]{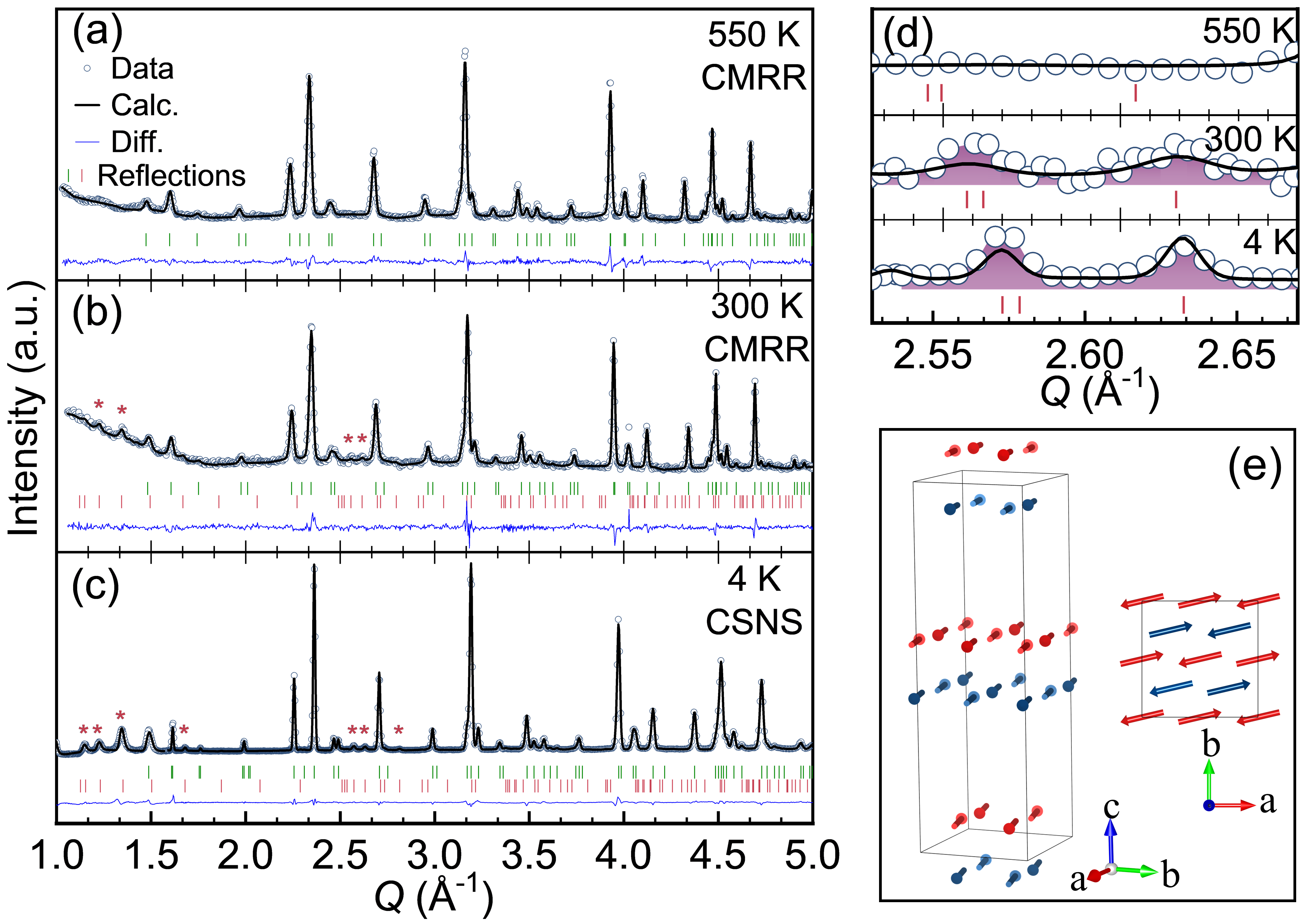}
\caption{Neutron diffraction patterns and magnetic structure of $\mathrm{Sr_3Fe_2O_5Br_2}$. Neutron diffraction patterns of $\mathrm{Sr_3Fe_2O_5Br_2}$ collected at (a) 550~K, (b) 300~K (CMRR), and (c) 4~K (CSNS). $*$ indicates magnetic diffraction peaks indexed by the propagation vector $(1/2, 1/2, 0)$. The upper tick marks (green) indicate the positions of the nuclear reflections, while the lower tick marks (red) denote the magnetic reflections. (d) Enlarged local neutron diffraction patterns, where {\protect\textbar} (red) represent the magnetic Bragg positions. (e) G-type magnetic structure model of $\mathrm{Sr_3Fe_2O_5Br_2}$, exhibiting antiparallel spin alignments both intra- and inter-layer. Here the black lines indicate enlarged magnetic (mag) cells with respect to nuclear (nuc) unit cells: $a_{\mathrm{mag}} = 2a_{\mathrm{nuc}}$, $b_{\mathrm{mag}} = 2b_{\mathrm{nuc}}$, and $c_{\mathrm{mag}} = c_{\mathrm{nuc}}$.}
\label{fig4}
\end{figure*}

Having established the macroscopic AFM orders in $\mathrm{Sr_3Fe_2O_5\textit{X}_2}$, temperature-dependent neutron powder diffraction measurements were performed on polycrystalline $\mathrm{Sr_3Fe_2O_5Br_2}$ samples, in order to probe the microscopic magnetic properties. Figures~\ref{fig4}(a)-(c) display the neutron diffraction patterns collected at select temperatures (4~K, 300~K and 550~K). Alongside nuclear Bragg reflections, additional magnetic Bragg peaks (marked by $*$ symbols) are clearly observed at 4~K and 300~K. Their peak intensities decrease upon warming, completely vanishing at 550~K (Figure~\ref{fig4}(d)), aligning with $T_{\mathrm{N}}$ inferred from the susceptibility minimum point. These magnetic reflections can be indexed with a magnetic propagation vector of $\boldsymbol{k} = (1/2, 1/2, 0)$. Magnetic symmetry analysis based on the parent crystallographic space group $I4/mmm$ allows the magnetic moments to be aligned either along the lattice $c$ axis or within the $ab$-plane. However, M\"ossbauer spectroscopy measurements~\cite{J.Solid.State.Chem.(1991)Ackerman} rule out the possibility of the $c$-axis spin configuration. Following the method proposed by Knee \textit{et al.}~\cite{Solid.State.Sci.(2004)Knee}, our refined G-type antiferromagnetic structure is illustrated in Figure~\ref{fig4}(e), where the spins are coupled antiferromagnetically both within the $ab$-plane and between adjacent layers. The in-plane magnetic moments are canted approximately $13^\circ$ from the [1 0 0] direction, similar to previous reports ~\cite{PRM(2025)May.AF, Solid.State.Sci.(2004)Knee}. Also, the magnitude of the ordered magnetic moment was refined at various temperatures, yielding approximately 4.08~$\mu_{\mathrm{B}}$/Fe at 4~K.

\subsection{Band gap}\label{subsec2}
\begin{figure*}[t]
\centering
\includegraphics[width=0.7 \textwidth]{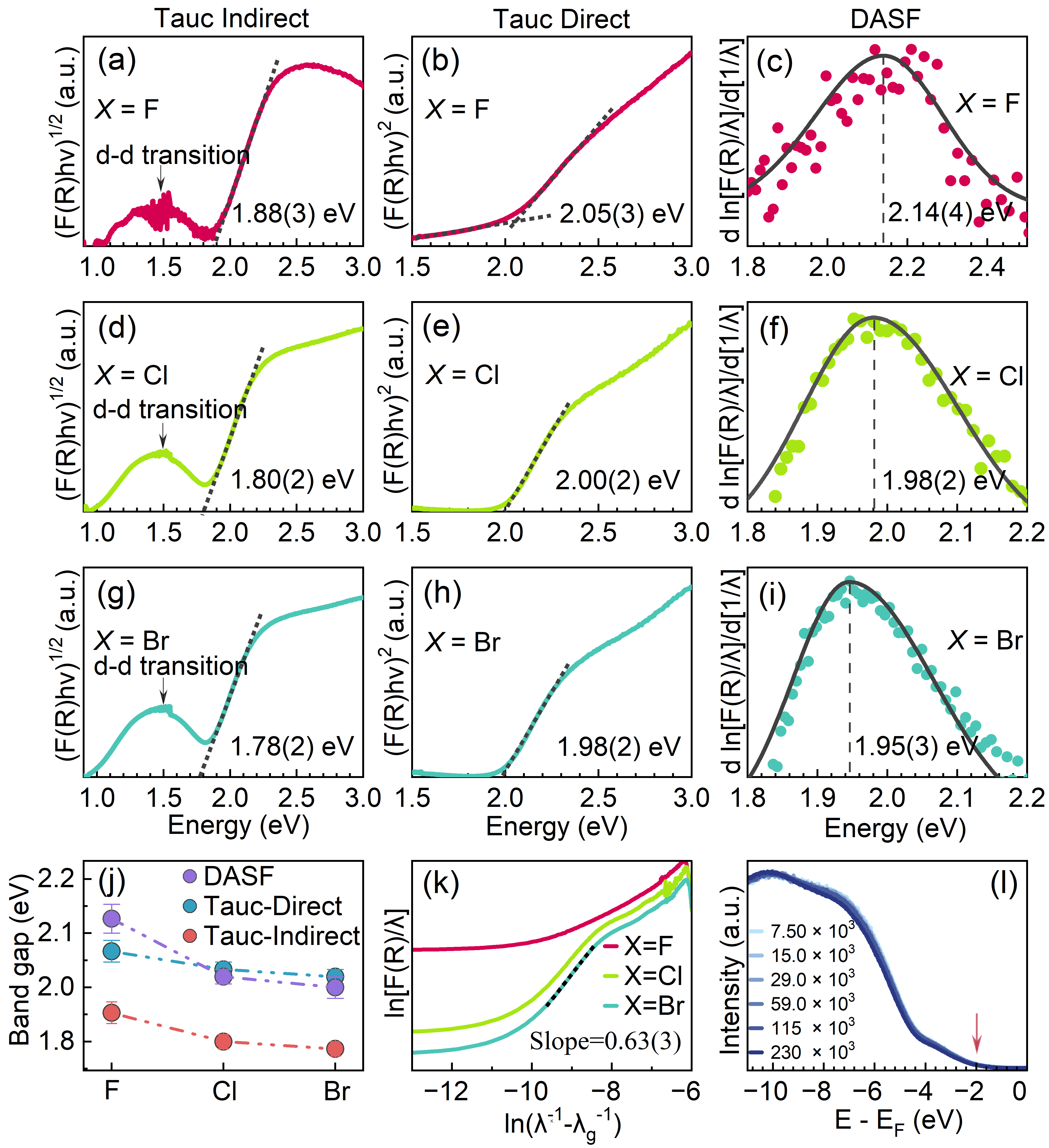}
\caption{Bandgap characterizations of $\mathrm{Sr_3Fe_2O_5\textit{X}_2}$. (a)-(i) Comparison of the optical bandgaps determined using Tauc plots (for direct and indirect transition models) and the DASF method. The determination of $E_{\mathrm{g}}$ is shown by the dash lines; (j) Optical bandgap versus halide species \textit{X}; (k) Transition indices derived from the DASF method; the index values for the F, Cl, and Br systems are 0.29, 0.58, and 0.63, respectively; (l) Angle-resolved photoemission spectroscopy (ARPES) measurements of the $\mathrm{Sr_3Fe_2O_5Br_2}$ single crystal, where the integrated density of states are plotted against energy using different photoelectron count rates (1/$\mathrm{s}$). An electronic band gap $\sim 2$~eV below the Fermi energy $E_{\mathrm{F}}$ is indicated by the red arrow.}
\label{fig5}
\end{figure*}

The semiconducting nature of $\mathrm{Sr_3Fe_2O_5\textit{X}_2}$ is verified by measuring both optical and electronic gaps (Figure~\ref{fig5} and Figure S4). The direct and indirect optical bandgaps of the three compounds were estimated via Tauc plot analysis ~\cite{Phys.Status.Solidi.B(1966)Tauc, JPCL(2023)zhong, JPCL(2018)Makula}, which was performed on their diffuse reflectance spectra. Detailed methodology is provided in the Supplementary Information Note 3. Because the Tauc method relies on a pre-assumed electronic transition type, the derivation of absorption spectrum fitting  method (DASF) ~\cite{Appli.Phy.B(2015)Souri, JPCC(2016)kirsch} was adopted, assigning the peak position of the derivative curve as the bandgap value. As shown in Figure~\ref{fig5}(j), the direct bandgap values extracted via the Tauc method are in agreement with the DASF results. The maximum transition index $\gamma$ (the slope of the curve) derived from the DASF method is approximately 0.63 (Figure~\ref{fig5}(k)), which is close to the theoretical index of 0.5 for a direct bandgap (2 for indirect bandgap). The electronic bandgap was measured by the ARPES study on $\mathrm{Sr_3Fe_2O_5Br_2}$ single crystals. The integrated intensity of density of states (DOS) is plotted against energy $E$ with respect to the Fermi energy $E_{\mathrm{F}}$. To exclude a space charge effect, different photoelectron count rates are used. It is evident that the DOS is only visible deeply below $E_{\mathrm{F}}$, yielding an electronic gap value greater than 2~eV (Figure~\ref{fig5}(l)). This number is comparable to the optical bandgaps in $\mathrm{Sr_3Fe_2O_5\textit{X}_2}$. Overall, the bandgap value of $\mathrm{Sr_3Fe_2O_5\textit{X}_2}$ displays a small decrease from \textit{X} = F, to \textit{X} = Cl and Br. This observation is likely because the halide species primarily modifies the local crystal-field environment rather than fundamentally altering the electronic band structure.

\subsection{Density functional theory (DFT) calculations}\label{subsec2}
\begin{figure*}[t]
\centering
\includegraphics[width=0.8\textwidth]{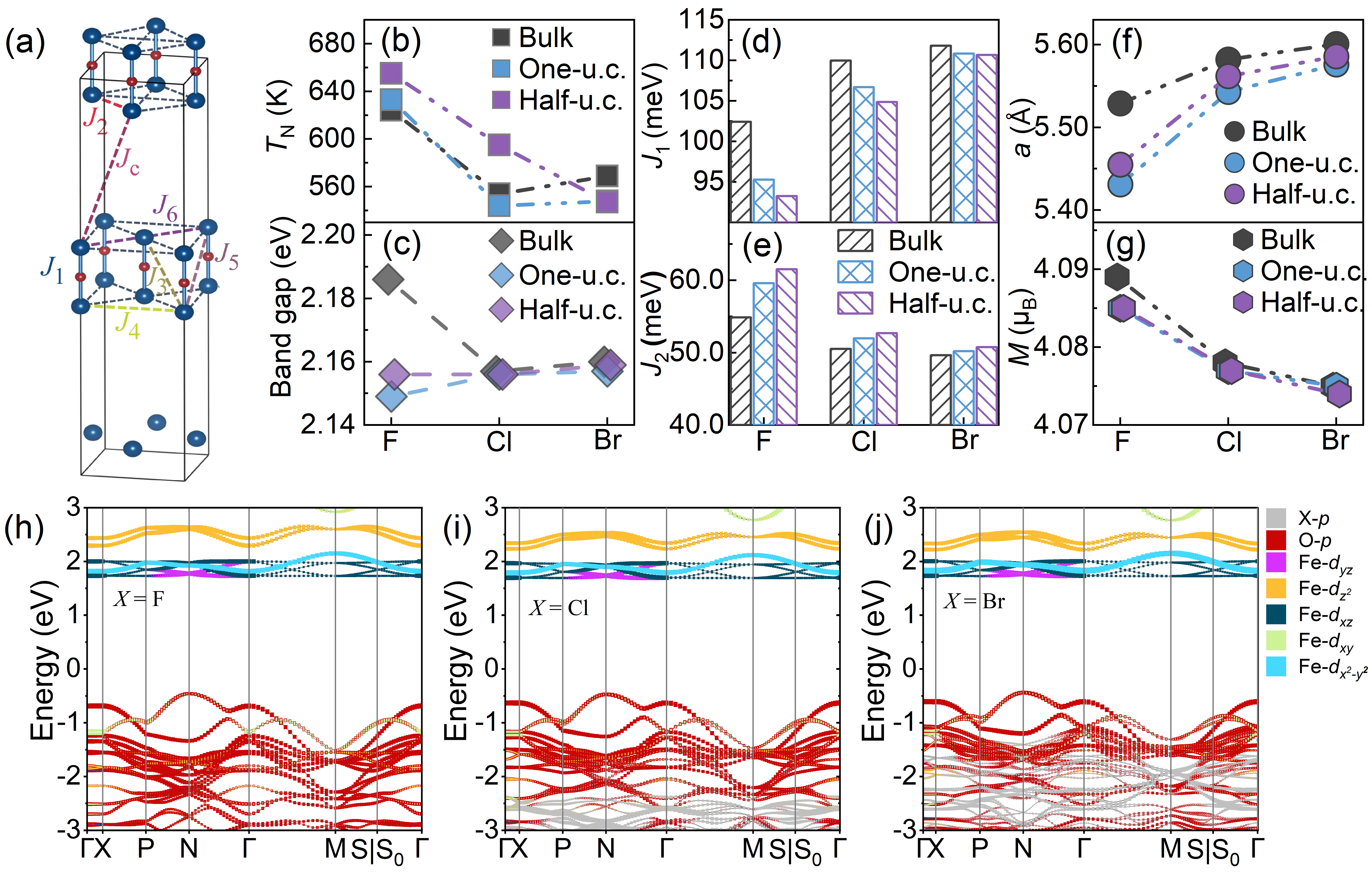}
\caption{Theoretical calculation results of $\mathrm{Sr_3Fe_2O_5\textit{X}_2}$. (a) Schematic diagram of the magnetic unit cell and the Heisenberg exchange coupling model with the main considered spin exchange constants $J_1$ to $J_6$, and $J_c$ ($a_{\mathrm{mag}} = \sqrt{2}a_{\mathrm{nuc}}$, $b_{\mathrm{mag}} = \sqrt{2}b_{\mathrm{nuc}}$, $c_{\mathrm{mag}} = c_{\mathrm{nuc}}$). Calculated (b) Néel temperature; (c) bandgaps; (d)-(e) Heisenberg exchange coupling constants $J_1$ and $J_2$, respectively; (f) lattice parameter $a$; and (g) magnetic moments. (h)-(j) The projected energy band structure with orbital weights for the three compounds of $\mathrm{Sr_3Fe_2O_5\textit{X}_2}$: (h) \textit{X} = F; (i) \textit{X} = Cl; (j) \textit{X} = Br, respectively.}
\label{fig6}
\end{figure*}

To understand the electronic and magnetic properties of $\mathrm{Sr_3Fe_2O_5\textit{X}_2}$, theoretical calculations based on the density functional theory (DFT) via the VASP software package are performed, as shown in Figure~\ref{fig6} and Supplementary Information Note 4. Based on the Heisenberg model (Figure~\ref{fig6}(a)), the out-of-plane $180^\circ$ Fe-O1-Fe superexchange coupling constant ($J_1$) within the bilayer motifs and the in-plane Fe-O2-Fe superexchange interaction ($J_2$) are dominating (Figures~\ref{fig6}(d)-(e), Table~\ref{tab2}). Upon halogen substitution from F to Br, $J_1$ increases from 102.41~meV, to 109.97~meV and 111.81~meV; while $J_2$ decreases from 54.88~meV to 50.52 and 49.66~meV. This is a bit surprising, since both the experimentally determined Fe-O1-Fe bond length $d_{\mathrm{Fe-O1-Fe}}$ and O2-Fe-O2 angle $\theta_{\mathrm{O2-Fe-O2}}$ do not evolve monotonically. On the other hand, the $J_1$ and $J_2$ numbers are much larger than other spin exchange coupling constants $J_3$ to $J_6$, particularly the inter-bilayer coupling ($J_c$), which is less than 1~meV. Resultingly, theoretical estimated Néel temperatures of bulk $\mathrm{Sr_3Fe_2O_5\textit{X}_2}$ display a decreasing trend from 624~K, 554~K, to 569~K, for \textit{X} = F to \textit{X} = Br and \textit{X} = Cl, respectively (Figure~\ref{fig6}(b)). This decreasing trend of $T_{\mathrm{N}}$, as well as other calculated quantities, such as bandgap magnitude ($\sim 2.16$~eV), lattice parameters and spin moment size ($\sim 4~\mu_{\mathrm{B}}$/Fe) are all consistent with the experimental numbers (Figures~\ref{fig6}(b),(c),(f),(g)).

To examine the stability of the AFM order in the 2D limit, DFT calculations are performed on $\mathrm{Sr_3Fe_2O_5\textit{X}_2}$ from bulk to one unit cell (one-u.c.) and half unit cell (half-u.c.) thickness. As indicated from Figures~\ref{fig6}(b)-(g), nearly all quantities display small variations, even when the sample thickness is reduced from bulk to the 2D limit. Particularly, because the intrinsic inter-bilayer coupling ($J_c$) is extremely weak, exfoliation does not disrupt the primary spin exchange interactions ($J_1$ and $J_2$). Therefore, the $T_{\mathrm{N}}$ of $\mathrm{Sr_3Fe_2O_5\textit{X}_2}$ are maintained down to the 2D limit. More interestingly, the $T_{\mathrm{N}}$ of $\mathrm{Sr_3Fe_2O_5F_2}$ and $\mathrm{Sr_3Fe_2O_5Br_2}$ display a further increase, to the highest number of 655~K for monolayer $\mathrm{Sr_3Fe_2O_5F_2}$ ($T_{\mathrm{N}} \sim 595$~K for monolayer $\mathrm{Sr_3Fe_2O_5Cl_2}$) (Figure~\ref{fig6}(b)), base on DFT calculations.

The orbitally projected band structures of $\mathrm{Sr_3Fe_2O_5\textit{X}_2}$ are shown in Figures~\ref{fig6}(h)-(j). The valence band maximum (VBM) is primarily derived from the O-$p$ orbitals, whereas the conduction band minimum (CBM) mainly originates from the Fe-$d$ orbitals. Obviously, the halogen $p$ orbitals are deeply buried below the Fermi level. This explains the nearly constant $\sim 2.1$~eV magnitude of the electronic bandgap in $\mathrm{Sr_3Fe_2O_5\textit{X}_2}$. Based on the DFT-calculated electronic band structures, an indirect electronic bandgap is inferred, contrasting with the conclusions derived from the optical spectra. However, the energy difference between the indirect transition ($N \rightarrow P$ point) and the direct transition ($N \rightarrow N$ point) in this system is exceptionally small (only approximately 0.04~eV).

\section{Discussion}\label{sec12}
The $\mathrm{Sr_3Fe_2O_5\textit{X}_2}$ compounds are derived from $\mathrm{Sr_3Fe_2O_7}$ ($T_{\mathrm{N}} = 115$~K)~\cite{PRB(2013)Peets} via substituting the apical $\mathrm{O^{2-}}$ site at the axial terminal of the $\mathrm{FeO_5}$ square pyramid with halide ions $\mathrm{\textit{X}^-}$ (\textit{X} = F, Cl, Br). This substitution reduces the planarity of the Fe-O2 planes and concurrently deviates the in-plane Fe-O2-Fe bond angles (equal to $\theta_{\mathrm{O2-Fe-O2}}$) from $\sim 176^\circ$ in $\mathrm{Sr_3Fe_2O_7}$, resulting in highly distorted $\mathrm{FeO_5\textit{X}}$ (\textit{X} = F, Cl, Br) octahedra. Despite these local structural distortions, the overall tetragonal symmetry (space group $I4/mmm$) is preserved. Importantly, with such substitution in $\mathrm{Sr_3Fe_2O_5\textit{X}_2}$, adjacent layers of halide atoms with vdW spatial gaps between the structural block motif are formed (Figure~\ref{fig1}(a)), forming layered structures. On the other hand, the asymmetric mixed-anion coordination environment induces a strong elongation of the local octahedra along the $z$-axis, leading to crystal-field splitting of the degenerate $d$ orbitals of the Fe ions. This non-centrosymmetric coordination environment relaxes the Laporte rule~\cite{JOSA(1925)laporte}, allowing light absorption via $d$-$d$ transitions, which are noticeable and have an energy scale $\sim 1.5$~eV, as shown in Figures~\ref{fig6}(a),~\ref{fig6}(d),~\ref{fig6}(g).

Although the halide ions of $\mathrm{Sr_3Fe_2O_5\textit{X}_2}$ do not directly participate in the magnetic exchange pathways, their replacements of the apical oxygen in $\mathrm{Sr_3Fe_2O_7}$ fundamentally alter the microscopic magnetic properties. Firstly, due to charge balance, the average valence state of iron is an unstable $\mathrm{Fe^{4+}}$ in $\mathrm{Sr_3Fe_2O_7}$, which displays a charge disproportionation at $\sim 343$~K ($2\mathrm{Fe^{4+}} \to \mathrm{Fe^{3+}} + \mathrm{Fe^{5+}}$)~\cite{J.Phy.Soci.Japan(2000)kuzushita, PRB(2013)Peets, PRL(2014)kim}. This phenomenon disappears in $\mathrm{Sr_3Fe_2O_5\textit{X}_2}$, because $\mathrm{Fe^{3+}}$ is a stable state. Consequently, the electronic structures of $\mathrm{Sr_3Fe_2O_5\textit{X}_2}$ are relatively simple, maintaining a nearly 2.1~eV gap. While DFT calculations indicate an indirect bandgap, direct transition of optical excitation might dominate the absorption edge due to its higher transition probability. This leads to the appeared discrepancy between the theoretical calculations and the optical spectra. More sensitive experimental techniques, such as photothermal deflection spectroscopy, are required to determine the transition mechanism.

Secondly, in $\mathrm{Sr_3Fe_2O_5\textit{X}_2}$, the half-filled $3d^5$ shell of $\mathrm{Fe^{3+}}$ with a high spin state $S = 5/2$ enables strong superexchange couplings through oxygen ligands. Based on DFT calculations, both out-of-plane (Fe-O1-Fe, $J_1$) and in-plane (Fe-O2-Fe, $J_2$) nearest-neighbor superexchange couplings are cooperating synergistically, which is directly responsible for the enormous five-fold increase in N\'eel temperature, from $T_{\mathrm{N}} \sim 115$~K in $\mathrm{Sr_3Fe_2O_7}$ to $T_{\mathrm{N}} \sim 580$~K in $\mathrm{Sr_3Fe_2O_5F_2}$. Microscopically, this strong superexchange arises from the pronounced hybridization between the \(\text{Fe } 3d\) and \(\text{O } 2p\) orbitals, which is facilitated by the ideal \(180^\circ\) \(\text{Fe}-\text{O}_1-\text{Fe}\) bond angle for \(J_1\) and the large \(\text{Fe}-\text{O}_2-\text{Fe}\) bond angle \((\sim 160^\circ\text{--}167^\circ)\) for \(J_2\). Such bonding geometries maximize the effective electron hopping integral,particularly the out-of-plane direction, yielding strong antiferromagnetic coupling. This synergy of magnetic couplings stabilizes a G-type AFM order with every $\mathrm{Fe^{3+}}$ spin antiparallel to all its nearest neighbors in $\mathrm{Sr_3Fe_2O_5\textit{X}_2}$. Although the calculated further-neighbor exchange interactions ($J_3$ to $J_6$, and $J_c$) are relatively weak, they still contribute to the large $T_{\mathrm{N}}$ in $\mathrm{Sr_3Fe_2O_5\textit{X}_2}$. Consequently, even though calculated $J_1 = 102.41$~meV for bulk $\mathrm{Sr_3Fe_2O_5\textit{X}_2}$ (\textit{X} = F) is smallest compared to that of 109.97~meV for \textit{X} = Cl and 111.81~meV for \textit{X} = Br, its calculated $T_{\mathrm{N}}$ is highest for \textit{X} = F. It should be noted that $J_c$ is negligibly small compared to all other spin exchange constants. As a result, reducing the thickness of $\mathrm{Sr_3Fe_2O_5\textit{X}_2}$ down to the 2D limit, the $T_{\mathrm{N}}$ still remains $\sim 600$~K or becomes even higher after relaxing the spin exchange strengths, according to DFT calculations. It would be very interesting to experimentally monitor and verify this predicted evolution of $T_{\mathrm{N}}$ in $\mathrm{Sr_3Fe_2O_5\textit{X}_2}$ as a function of flake thickness down to the true 2D limit.

\section{Conclusion}\label{sec13}
In conclusion, we promote a layered vdW material family with beyond-room-temperature antiferromagnetic orders in $\mathrm{Sr_3Fe_2O_5\textit{X}_2}$. The layered form of $\mathrm{Sr_3Fe_2O_5F_2}$ displays the largest $T_{\mathrm{N}} \sim 580$~K. Single crystals of $\mathrm{Sr_3Fe_2O_5Cl_2}$ with lateral dimension reaching more than 20~mm are obtained; while one-unit-cell thick thin-flakes of $\mathrm{Sr_3Fe_2O_5Br_2}$ are readily isolated via mechanical exfoliation, highlighting the layered nature of $\mathrm{Sr_3Fe_2O_5\textit{X}_2}$. DFT calculations demonstrate that the nearest neighbor out-of-plane and in-plane superexchange interactions ($J_1$ and $J_2$) within the bilayer motifs of $\mathrm{Sr_3Fe_2O_5\textit{X}_2}$ are mainly responsible for the large $T_{\mathrm{N}}$, which is predicted to persist even upon exfoliation down to the monolayer limit. Our work highlights that bilayer RP-phase iron-based oxyhalides $\mathrm{Sr_3Fe_2O_5\textit{X}_2}$ represent a novel class of van der Waals magnets with approaching $\sim 600$~K AFM magnetic order. Its high $T_{\mathrm{N}}$ far above room temperature and cleavable, mica-like nature, as well as the wide optical bandgap and semiconducting nature, offer an ideal platform for exploring magnetism in the two-dimensional limit, promising for spintronic devices and magnetic memory applications.

\section{Materials and method}
\subsection{Synthesis}\label{subsec2}
Polycrystalline $\mathrm{Sr_3Fe_2O_5F_2}$ samples were prepared via chemical fluorination ~\cite{J.Solid.State.Chem(2012)Hancock, Chem.Mat(2011)Tsujimoto}. The $\mathrm{Sr_3Fe_2O_7}$ precursor was produced via a typical solid-state reaction method~\cite{Angew.Chem(2008)Kageyama}. Fluorination was carried out in an evacuated quartz tube. Specifically, $\mathrm{Sr_3Fe_2O_7}$ and PVDF were mixed in a 1:1.5 molar ratio, ground together, and vacuum-sealed in the tube. The mixture was heated at 350~$^\circ$C for 24~h, yielding a reddish-brown powder. Polycrystalline $\mathrm{Sr_3Fe_2O_5Cl_2}$ and $\mathrm{Sr_3Fe_2O_5Br_2}$ were synthesized using a similar solid-state approach as previously reported ~\cite{Solid.State.Sci.(2004)Knee}. Stoichiometric amounts of SrO (decompose from $\mathrm{SrCO_3}$ at 1100~$^\circ$C, 4 days), $\mathrm{Fe_2O_3}$ (dried at 200~$^\circ$C for 2~h) and $\mathrm{SrCl_2}$ ($\mathrm{SrBr_2}$) were mixed and ground thoroughly in an agate mortar within a glove box. The resulting mixture was pelletized and placed in an alumina crucible, before being vacuum-sealed in a quartz ampoule. Subsequently, the tube was sintered at 850~$^\circ$C for 48~h to obtain a red product.

Single crystals of $\mathrm{Sr_3Fe_2O_5Cl_2}$ and $\mathrm{Sr_3Fe_2O_5Br_2}$ were grown using a high-temperature flux method. $\mathrm{Sr_3Fe_2O_5Cl_2}$ was prepared following the protocol described in Ref.~\cite{PRM(2025)May.AF}. Starting materials with a molar ratio of $\mathrm{SrCO_3:Fe_2O_3:SrCl_2}$ = 2:1:40 were mixed, ground, and loaded into an alumina crucible with a lid. The crucible was placed in a box furnace, heated to 1100~$^\circ$C, held for 24~h, and then cooled to 875~$^\circ$C at a rate of 1~$^\circ$C/h. $\mathrm{Sr_3Fe_2O_5Cl_2}$ crystals were also obtained using the same heating profile with a molar ratio of $\mathrm{SrCO_3:Fe_2O_3:SrCl_2}$ = 2:1:15. A representative crystal is shown in Figure~\ref{fig2}(a). Attempts to grow $\mathrm{Sr_3Fe_2O_5Br_2}$ single crystals using this specific protocol failed. Instead, $\mathrm{Sr_3Fe_2O_5Br_2}$ single crystals were successfully grown using $\mathrm{SrCO_3}$, $\mathrm{FeC_2O_6H_4}$, and $\mathrm{SrBr_2}$ (from $\mathrm{SrBr_2 \cdot 6H_2O}$ at 345~$^\circ$C, 1 day), as starting materials in a molar ratio of 2:2:5. The loaded crucible was sintered at 850~$^\circ$C for 5~h in a Czochralski furnace, followed by a slow cooling process to 600~$^\circ$C over 72~h before the furnace was switched off. Plate-like $\mathrm{Sr_3Fe_2O_5Br_2}$ single crystals with typical lateral dimensions of approximately 5~mm are shown in Figure~\ref{fig2}(a). Notably, repeated growth attempts using a box furnace failed to yield the target phase. We attribute this successful crystallization to the enhanced vertical thermal gradient characteristic of the Czochralski furnace. Single-crystalline $\mathrm{Sr_3Fe_2O_5F_2}$ was not obtained, likely due to the high chemical stability of $\mathrm{SrF_2}$. High-pressure synthesis techniques may facilitate the growth of this compound~\cite{Chem.Mat(2011)Tsujimoto}.

\subsection{Characterization}\label{subsec2}
Single-crystal X-ray diffraction (SCXRD) data were collected using a Bruker APEX-II CCD X-ray diffractometer at 150~K, while powder X-ray diffraction (PXRD) measurements were performed at room temperature using a Rigaku MiniFlex diffractometer. 

Temperature-dependent magnetic susceptibility measurements were performed on a Physical Property Measurement System (PPMS) from 300 to 700~K under a magnetic field of 70~kOe. Isothermal magnetization curves were recorded using a Magnetic Property Measurement System (MPMS). 

Neutron diffraction data for the polycrystalline $\mathrm{Sr_3Fe_2O_5Br_2}$ samples were collected at various temperatures. High-temperature neutron powder diffraction measurements were performed at the China Mianyang Research Reactor (CMRR, China), using a neutron wavelength of $\lambda = 1.8848$~\AA{}, and diffraction patterns were recorded at 300~K and 550~K, respectively. Low-temperature diffraction data at 4~K were collected using the General Purpose Powder Diffractometer (31113.02.CSNS.GPPD) at China Spallation Neutron Source (CSNS, China). Rietveld refinements of the diffraction data were conducted using the Fullprof suite~\cite{PHYSICA.B(1993)Rodríguez}, and magnetic structures were determined using SARAh~\cite{Phys.B.Con.Mat.(2000)Will}.

The Ultraviolet-Visible (UV-Vis) diffuse reflectance spectra of all polycrystalline samples were measured using a Cary 5000 spectrophotometer (Agilent) in the spectral range of 300$\sim$1500~nm at room temperature.

ARPES measurements were performed using a home-lab ARPES system equipped with a DA30L electron analyzer and a helium lamp light source with 21.2 eV photon energy. The angle resolution was set to 0.3 degrees and the energy resolution was set to 20 meV.

TEM specimens were prepared using a focused ion beam (FIB) instrument (Thermo Fisher Helios 5CX) equipped with a Ga ion source. The specimens were sequentially thinned down to 20~nm using decreasing accelerating voltages, followed by a final low-voltage (1~kV) ion milling step to remove residual amorphous layers. Atomic-resolution structural characterization was performed using a spherical aberration-corrected transmission electron microscope (JEM NeoARM200, JEOL Ltd., Tokyo) operated at 200~kV. The dataset was acquired using a high-speed direct electron detection camera (DECTRIS ARINA), yielding a four-dimensional STEM with data volume of $192 \times 192 \times 400 \times 400$.

First-principles calculations were performed using the Vienna Ab initio Simulation Package (VASP)~\cite{93prb,96prb} with the projector augmented-wave (PAW) method and a plane-wave cutoff energy of 450~eV. The Perdew--Burke--Ernzerhof (PBE) functional within the generalized gradient approximation (GGA) was employed~\cite{96prl}. The DFT + $U$ method~\cite{DFTU} was employed with an effective Hubbard parameter $U_{\mathrm{eff}} = 4$~eV for Fe $3d$ electrons.
For bulk systems, the experimental lattice constants were adopted. For two-dimensional (2D) systems, the in-plane lattice constants were fully optimized. All atomic positions were relaxed until the Hellmann--Feynman forces on each atom were below 0.01~eV/\AA{}. The van der Waals interactions were taken into account using the DFT-D3 correction scheme~\cite{DFT-D3}. Brillouin-zone integrations were carried out using $\Gamma$-centered Monkhorst--Pack $k$-point meshes~\cite{MK}. For the magnetic unit cell corresponding to a $\sqrt{2}\times\sqrt{2}\times1$ supercell of the primitive cell, $6\times6\times2$ and $6\times6\times1$ $k$-point grids were employed for bulk and 2D systems, respectively.

\medbreak

\clearpage
\section{References}
\bibliographystyle{apsrev4-2}
\bibliography{SFOX3252}

\medbreak
\textbf{Acknowledgments} 
Work at SYSU was supported by the National Natural Science Foundation of China (Grants No. 12574140, No. 12494591, No. 12425404, U25A20193, 92565303), the National Key Research and Development Program of China (Grant No. 2023YFA1406500), the CAS Superconducting Research Project (Grant No. SCZX-0101), the Guangdong Provincial Key Laboratory of Magnetoelectric Physics and Devices (Grant No. 2022B1212010008), Research Center for Magnetoelectric Physics of Guangdong Province (Grant No. 2024B0303390001), and the Fundamental Research Funds for the Central Universities, Sun Yat-sen University (Grant No. 74130-31610055). We thank the staff members of the General Purpose Powder Diffractometer (31113.02.CSNS.GPPD) at the China Spallation Neutron Source (CSNS), for providing technical support and assistance in data collection and analysis.
\medbreak
\medbreak
\textbf{Competing interests} The authors declare that they have no conflict of interest.
\medbreak
\textbf{Additional information} Supplementary information available. 

\clearpage
\section{Tables}\label{sec5}

\begin{table}[h]
\centering
\caption{Refined lattice parameters, select bond lengths (in units of~\AA{}) and bond angles (in units of $^\circ$) from powder x-ray diffraction patterns of $\mathrm{Sr_3Fe_2O_5\textit{X}_2}$ at room temperature.}
\footnotesize
\begin{tabular*}{\linewidth}{@{\extracolsep{\fill}}cccc@{}}
\hline
 &Sr$_3$Fe$_2$O$_5$F$_2$ & Sr$_3$Fe$_2$O$_5$Cl$_2$  &  Sr$_3$Fe$_2$O$_5$Br$_2$ \\
\hline
$a$ (\AA)& 3.908(1)& 3.947(1)& 3.960(1)\\
     $c$ (\AA)& 21.383(1)& 23.799(2)& 25.425(1)\\
     vdW gap (\AA)& 1.652(1)& 2.132(1)& 2.479(1)\\
     Fe--O1 (\AA)&  1.911(3) &  1.889(2) & 1.922(3)  \\
     Fe--O2 (\AA)& 1.966(2)&  2.001(1)& 1.994(1)\\
     Fe--X (\AA)& 2.478(2)&  2.938(1)& 3.193(2)\\
     Sr1--O1 (\AA)&  2.764(1)&  2.791(1)& 2.800(1)\\
     Sr1--O2 (\AA)&  2.897(2)&  2.972(1)& 2.930(2)\\
     Sr2--O2 (\AA)& 2.579(2)&  2.485(1)& 2.543(1)\\
     Sr2--X (\AA)&  2.823(2)&  2.998(1)& 3.113(1)\\
    Fe--O1--Fe ($^{\circ}$)& 180.00&  180.00& 180.00\\
    O2--Fe--O2 ($^{\circ}$)&  167.18(1)&  160.82(1)& 166.27(1)\\
    O1--Sr1--O1 ($^{\circ}$)& 180.00&  180.00& 180.00\\
    O2--Sr1--O2 ($^{\circ}$)& 84.83(1)&  83.19(2)& 84.99(2)\\

\hline
\end{tabular*}
\label{tab1}
\end{table}

\begin{table*}[]
\centering
    \caption{Exchange parameters $J$ (meV) and N\'eel temperatures $T_{\mathrm{N}}$ (K) obtained from DFT calculations.}
    \label{tab:combined_data}
    \begin{tabular}{clcccccccc}
        \hline
         & {Materials} & $J_1$ & $J_2$ & $J_3$ & $J_4$ & $J_5$ & $J_6$ & $J_c$ & $T_{\mathrm{N}}$ \\
        \hline
        \multirow{3}{*}{Bulk}
        & $\mathrm{Sr_3Fe_2O_5F_2}$  & 102.41 & 54.88 & 3.31 & 2.34 & 0.49  & 1.20 & 0.09  & 624  \\
        & $\mathrm{Sr_3Fe_2O_5Cl_2}$ & 109.97 & 50.52 & 3.92 & 2.97 & -0.08 & 1.28 & -0.04 & 554  \\
        & $\mathrm{Sr_3Fe_2O_5Br_2}$ & 111.81 & 49.66 & 3.48 & 2.37 & 0.48  & 1.05 & 0.08  & 569 \\
        \hline
         \multirow{3}{*}{One unit-cell} 
        & $\mathrm{Sr_3Fe_2O_5F_2}$  & 95.28  & 59.60 & 3.25 & 2.48 & 0.31  & 1.46 & 0.67  & 633  \\
        & $\mathrm{Sr_3Fe_2O_5Cl_2}$ & 106.70 & 51.98 & 3.58 & 2.94 & -0.12 & 1.40 & -0.47 & 544  \\
        & $\mathrm{Sr_3Fe_2O_5Br_2}$ & 110.82 & 50.19 & 3.48 & 2.07 & 0.27  & 1.21 & 0.58  & 548  \\
         \hline
         \multirow{3}{*}{Half unit-cell} 
        & $\mathrm{Sr_3Fe_2O_5F_2}$  & 93.24  & 61.53 & 3.53 & 3.15 & 0.17  & 1.93 & --    & 655  \\
        & $\mathrm{Sr_3Fe_2O_5Cl_2}$ & 104.87 & 52.68 & 3.02 & 2.83 & -0.09 & 0.76 & --    & 595  \\
        & $\mathrm{Sr_3Fe_2O_5Br_2}$ & 110.67 & 50.77 & 3.50 & 2.68 & 0.19  & 1.65 & --    & 547  \\
\hline
    \end{tabular}
\label{tab2}
\end{table*}
\clearpage

\section{Figures}\label{sec6}

\clearpage

\end{document}